# Bottleneck accumulation of hybrid magneto-elastic bosons

Dmytro A. Bozhko,[1,2] Peter Clausen,[1] Gennadii A. Melkov,[3] Victor S. L'vov,[4] Anna Pomyalov,[4]
Vitaliy I. Vasyuchka,[1] Andrii V. Chumak,[1] Burkard Hillebrands,[1] and Alexander A. Serga[1]

[1]*Fachbereich Physik and Landesforschungszentrum OPTIMAS, Technische Universität Kaiserslautern, 67663 Kaiserslautern, Germany*
[2]*Graduate School Materials Science in Mainz, Kaiserslautern 67663, Germany*
[3]*Faculty of Radiophysics, Electronics and Computer Systems, Taras Shevchenko National University of Kyiv, Kyiv 01601, Ukraine*
[4]*Department of Chemical Physics, Weizmann Institute of Science, Rehovot 76100, Israel*
(Dated: December 18, 2016)

It is known that an ensemble of magnons, quanta of spin waves, can be prepared as a Bose gas of weakly interacting quasiparticles with conservation of the particle number. Furthermore, the thermalization of the overpopulated magnon gas can lead to the formation of a Bose-Einstein condensate at the bottom of a spin-wave spectrum. However, magnon-phonon scattering processes can significantly modify this scenario and new quasiparticles are formed — magneto-elastic bosons. Our observations of a parametrically populated magnon gas in a single-crystal film of Yttrium Iron Garnet by means of wavevector-resolved Brillouin light scattering spectroscopy including magneto-elastic coupling resulted in the discovery of a novel condensation phenomenon: A spontaneous accumulation of hybrid magneto-elastic bosonic quasiparticles at the intersection of the lowest magnon mode and a transversal acoustic wave.

Macroscopic quantum states—Bose-Einstein condensates (BECs) can be created in overpopulated gases of bosonic quasiparticles (excitons [1], polaritons [2–4], magnons [5–8], photons [9], etc.). However, interactions between quasiparticles of different nature [10], for example between magnons and phonons in a magnetic medium [11–18], can significantly alter the properties of these gases and thus modify the condensation scenarios [19, 20]. Here, we report on the discovery of a novel condensation phenomenon mediated by the magnon-phonon interaction: A bottleneck accumulation of hybrid magneto-elastic bosons. We have found that the transfer of quasiparticles toward a BEC state is almost fully suppressed near the intersection point between the magnon and phonon spectral branches. Such a bottleneck leads to a strong spontaneous accumulation of the quasiparticles trapped near the semi-linear part of the magnon-phonon hybridization area. As opposed to BEC, which is a consequence of equilibrium Bose statistics, the bottleneck accumulation is determined by varying interparticle interactions. Furthermore, unlike BEC, the accumulated magneto-elastic bosons possess a nonzero group velocity, making them promising data carriers in prospective magnon spintronics [21] circuits.

The spectral characteristics of an overpopulated magnon-phonon gas were studied at room temperature in an Yttrium Iron Garnet (YIG, $Y_3Fe_5O_{12}$) film by frequency-, time- and wavevector-resolved Brillouin light scattering (BLS) spectroscopy (see Fig. 1a and Appendix). Following the approach, which was established in previous experiments on a magnon BEC [6–8], magnons were injected into the in-plane magnetized YIG film via parallel parametric pumping [22]. In this process, one photon of a pumping electromagnetic field of frequency $\omega_P$ splits into two magnons having opposite wavevectors $\pm q$ and frequency $\omega_P/2$ (see the red arrow in Fig. 1b). Single-crystal YIG is a ferrimagnetic insulator that combines the uniquely low magnon and phonon damping [23] with a pronounced nonlinear magnon dynamics thus allowing for an efficient thermalization of externally injected magnons. This thermalization causes an increase in the chemical potential of the pumped magnon gas, and when it becomes equal to the minimum magnon energy a magnon BEC forms at this spectral point (see Fig. 1b) [6, 7].

Surprisingly, by virtue of the wavevector-resolved BLS technique we have detected another spectral point where the quasiparticles accumulate away from the global energy minimum of a pure magnon spectrum (compare the orange and green dots in Fig. 1b). BLS intensity maps representing the population of the magnon-phonon spectrum are presented in Fig. 2a,b as a function of frequency and wavenumber $q_\parallel$ ($q \parallel H$) for two different bias magnetic fields and a relatively small pumping power $P_P$ of 2.6 W. In spite of the fact that the threshold for magnon BEC formation is still not reached at such power levels, one can see an intense population peak near the region of the hybridization between the magnon and the transversal acoustic phonon dispersion branches (see white lines in Fig. 2). As the bias magnetic field is shifted from 1265 Oe to 1735 Oe, the population peak shifts in frequency and wavenumber together with the magneto-elastic crossover region. At the same time, there are no peculiarities in the thermal spectrum, measured at the same conditions, but without the application of pumping (see Fig. 2c).

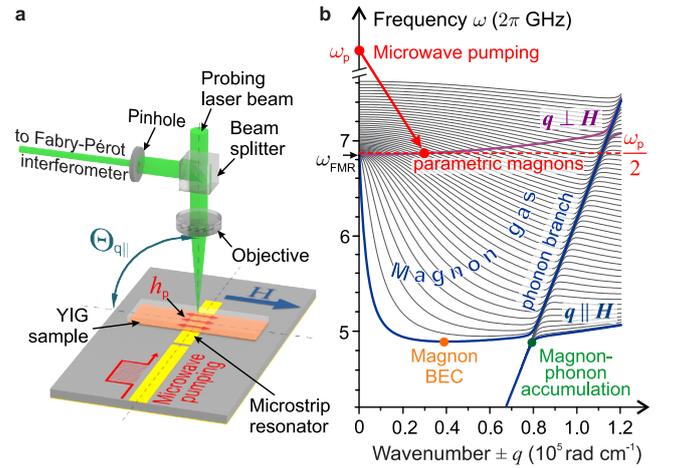

FIG. 1. Set-up for excitation and observation of magnons, phonons, and magneto-elastic bosons. **a**, Schematic illustration of the experimental set-up. The resonator concentrates the applied microwave energy and induces a pumping microwave Oersted field $h_P$ oriented along the bias magnetic field $H$, thus realizing conditions for the parallel parametric pumping. The probing laser beam is focused onto a YIG film placed on top of a microstrip resonator. The light inelastically scattered by magnons is redirected to a Fabry-Pérot interferometer for frequency and intensity analysis. Wavenumber-selective probing of magnons with wavevectors $\pm q \parallel H$ is realized by varying the incidence angle $\Theta_{q_\parallel}$ between the field $H$ and the probing laser beam. **b**, Magnon-phonon spectrum of a 6.7 $\mu$m-thick YIG film calculated for $H = 1735$ Oe. 47 thickness modes with $q \parallel H$. The upper thick curve shows the most effectively parametrically driven [24] lowest magnon mode with $q \perp H$. The arrow illustrates the magnon injection to the frequency $\omega_P/2$ slightly above the ferromagnetic resonance frequency $\omega_{FMR}$.

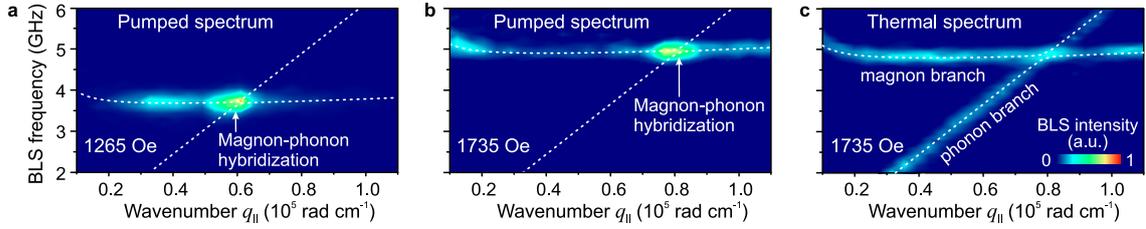

FIG. 2. Magnon-phonon spectra and their population under different pumping conditions. White dashed lines represent the calculated dispersion relation for the lowest magnon branch ($q \parallel H$), hybridized with a transversal acoustic mode. Strong population of the magnon-phonon hybridization region is clearly visible in the parametrically pumped spectra (**a-b**) measured in the case of a $500\,\mu$m–wide pumping area. The width of the population peak is of the order of the wavenumber resolution of the experimental set-up, i.e. $\pm 0.02 \cdot 10^5$ rad cm$^{-1}$. No distinct peculiarities in the thermal spectrum measured for zero pump power are detected (**c**).

It is important that the population peak is visible only if a sufficiently wide microstrip resonator is used to create the pumping microwave field $h_\mathrm{p}$ (see Fig. 1a). This then confirms that the accumulation occurs into the magnon-phonon hybridisation region, where quasiparticles possess rather high group velocities—the widening of the width of the microstrip reduces their leakage from the pumping area. Obviously, the accumulation mechanism must significantly differ from the conventional BEC as no energy minimum exists in this region.

To reveal characteristic features of the observed accumulation effect, we performed time-resolved BLS measurements for both the narrow ($50\,\mu$m) and the broad ($500\,\mu$m) pumping areas. The results are presented in Fig. 3. Evidently, in both cases there are some delays between the application of the pumping pulse and population of the lowest energy states of the spin-wave spectrum by thermalized magnons. The delay time consists of two components. The first thereof is the time required for the parametric process to develop. The second one is the time needed for the initially injected parametric magnons to scatter to the bottom of the spectrum, where magnons first appear in the exchange region of the spectrum around $q_\parallel = 1.1 \cdot 10^5$ rad cm$^{-1}$. Then, a pronounced population peak is formed at the magneto-elastic crossover at $q_\parallel = 0.8 \cdot 10^5$ rad cm$^{-1}$ in the case of the widen pumping area (see Fig. 3a). On the contrary, in compliance with our previously reported results [25], no accumulation is visible in this spectral region in the case of the narrow pumping area (see Fig. 3b). During further time evolution, the magnons occupy energy states around the global energy minimum $q_0 = 0.4 \cdot 10^5$ rad cm$^{-1}$ and tend to form a Bose-Einstein condensate. The enhancement of the BEC formation in a freely evolving magnon gas leads to the appearance of the BEC's density peak just after the termination of the pumping pulse [7].

Remarkably, the magneto-elastic peak emerges at pumping powers at least ten times smaller than the threshold of the BEC formation (see Fig. 3a). Nevertheless, with the increase of the pumping power (and consequently of the BEC density), further growth in the population of the magneto-elastic peak is suppressed (see black and red lines in the right panel of Fig. 3a).

To understand the observed phenomena, we should consider the peculiarities of the scattering processes leading to the thermalization of the parametrically-driven magnon gas. A microwave electromagnetic field with frequency $\omega_\mathrm{p}$ and wavenumber $q_\mathrm{p} \approx 0$ excites parametric magnons, distributed over the isofrequency surface $\omega_q^m = \omega_p/2$ (see Fig. 1b). Due to "2 ⇒ 2" four-magnon scattering, satisfying the laws of conservation $\omega_{q_1}^m + \omega_{q_2}^m = \omega_{q_3}^m + \omega_{q_4}^m$ and $q_1 + q_2 = q_3 + q_4$, the parametrically excited magnons redistribute over the entire $q$-space, such that the flux of the magnon number $n(q)$ is mostly directed toward lower frequencies. To find the fastest track in the $q$-space for magnons flowing from the pumping area toward the BEC destination point, we note, that at relatively high frequencies, the main contribution to the scattering amplitude $T_{q_1,q_2;q_3,q_4}$ is given by the exchange interactions for which $T_q \equiv T_{q,q;q,q} \propto q^2$. Therefore, the magnons with the largest wave numbers $q$ are most efficiently thermalized. For an in-plane magnetized YIG film, they belong to the lowest magnon mode with $q \parallel H$ [25, 26]. Thus, the fastest way for the parametric magnons from the pumping area toward the bottom of the energy well lies along the dispersion curve of this mode.

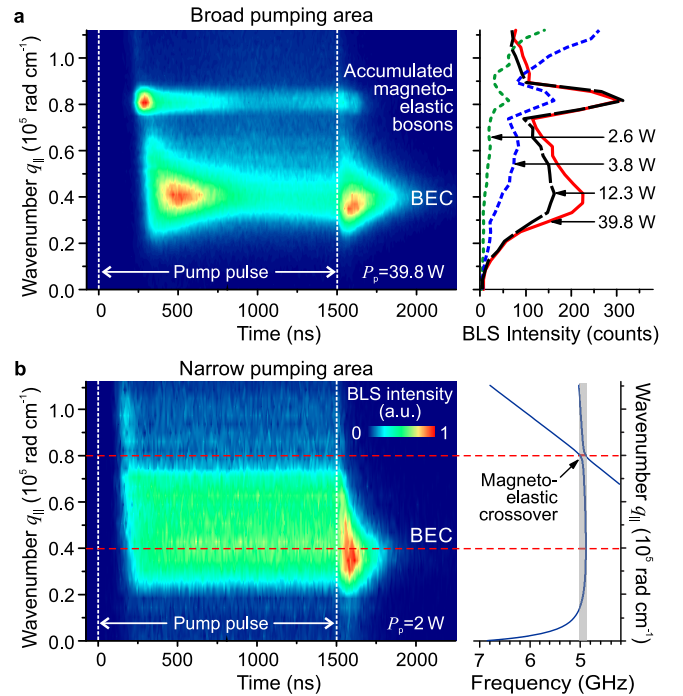

FIG. 3. Temporal dynamics of the magnon gas density. **a**, Pronounced accumulation of magneto-elastic bosons is evident for the broad pumping area of $500\,\mu$m width. The right panel shows the pumping power dependencies of the quasiparticle density integrated by time. The saturation of the magneto-elastic peak at high pumping powers argues in favour of a relationship between the magnon BEC and the accumulation phenomenon. No accumulation effect is visible in (**b**) due to a strong leakage of fast magneto-elastic bosons from the narrow pumping area of $50\,\mu$m width. The pumping power is significantly decreased due to the increase in the density of a microwave current in the narrow microstrip. In all measurements BLS data are collected in the frequency band of 150 MHz near the bottom of the magnon spectrum (see the shaded area in the right panel of (**b**)). The bias magnetic field is $H = 1735$ Oe.

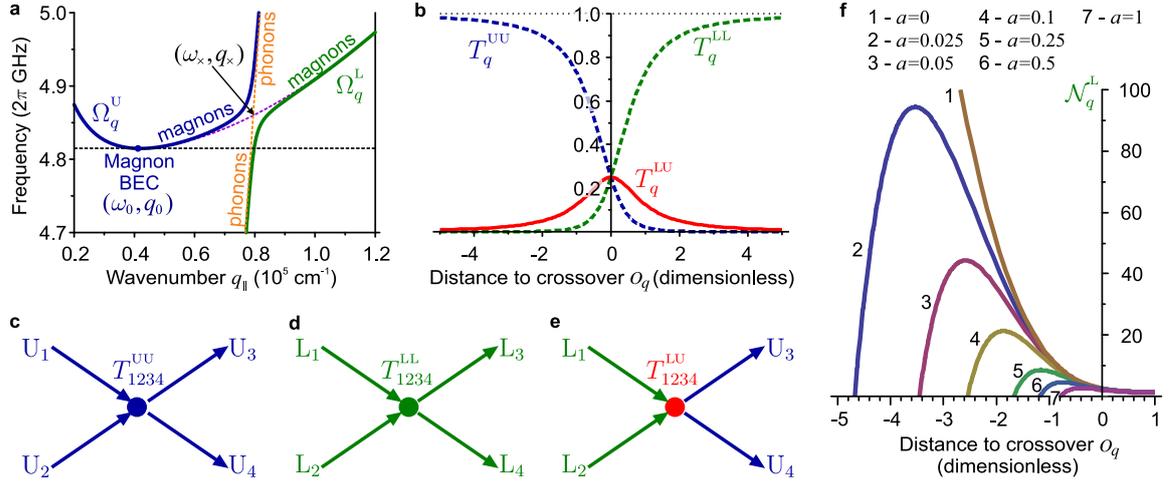

FIG. 4. Theory of the bottleneck accumulation. **a**, Calculated magnon-phonon spectrum near the hybridization region. **b**, Interaction amplitudes $T_q^{UU}$, $T_q^{LL}$ and $T_q^{UL}$, normalized by $T_q$, vs dimensionless distance from the hybridization crossover $o_q$. **c-e**, Schematic representation of the scattering and cross-scattering 4-particle processes in the hybridization region. **f**, Solutions of the dimensionless L-MEM density $\mathcal{N}_q^L$, for different values of the parameter $a$.

In the presence of magnon-phonon coupling, the unperturbed magnon $\omega_q^m$ and phonon $\omega_q^p$ spectra (see Fig. 4a, dashed lines) split up into Upper and Lower Magneto-Elastic Modes (U-MEM, $\Omega_q^U$ and L-MEM, $\Omega_q^L$), (Fig. 4a, solid lines). The population dynamics of these modes are determined by both the intramodal (2L⇔2L, 2U⇔2U) and the intermodal (2L⇔2U) scattering of hybrid magneto-elastic bosons with the scattering amplitudes $T_{q_1,q_2;q_3,q_4}^{LL}$, $T_{q_1,q_2;q_3,q_4}^{UU}$, and $T_{q_1,q_2;q_3,q_4}^{LU}$, respectively. These scattering amplitudes originate from the four-magnon scattering amplitude $T_{q_1,q_2;q_3,q_4}$ and are, therefore, proportional to the magnon contribution to MEMs. It can be shown (see Appendix) that this contribution is given by $\cos \varphi_q$ for the L-MEM and $\sin \varphi_q$ for the U-MEM, whose values are determined by the dimensionless frequency distance $o_q$ from the crossover:

$$\cos \varphi_q = \frac{1}{\sqrt{2}}\left[1 + \frac{o_q}{\sqrt{1+o_q^2}}\right]^{1/2}, \quad o_q = \frac{\omega_q^p - \omega_q^m}{\Delta}, \quad (1)$$

where $\Delta = \Omega_{q_\times}^U - \Omega_{q_\times}^L$ is the magneto-elastic frequency and $q_\times$ is the crossover wavevector, at which $\omega_{q_\times}^p = \omega_{q_\times}^m \equiv \omega_\times$ (Fig. 4a). In particular

$$T_q^{LL} = T_q \cos^4 \varphi_q, \quad T_q^{LU} = T_q \cos^2 \varphi_q \sin^2 \varphi_q, \quad (2)$$

and $T_q^{UU} \simeq T_q \sin^4 \varphi_q$. Plots of $T_q^{LL}$, $T_q^{UU}$, and $T_q^{LU}$ vs. $o_q$ are shown in Fig. 4b. As expected, for positive $o_q$ (i.e. $q > q_\times$), the L-MEM becomes a pure magnon mode with $T_q^{LL} \to T_q$, while the U-MEM becomes a pure phonon mode with $T_q^{UU} \to 0$ (see Figs. 4a and 4b). The intermodal amplitude $T_q^{LU}$ (shown by red solid line in Fig. 4b) requires the presence of the magnon contribution in both the L- and U-MEMs, therefore its value is significant only in the hybridization region, where $|o_q| \lesssim 1$.

To find the L-MEM population around the hybridization area, we consider the stationary balance equation for density of L-MEM quasiparticles:

$$\frac{d\mu_q^L}{dq} = F_q^{LU}, \quad (3)$$

Here, $\mu_q^L$ is the L-MEM flux towards the hybridization area, originating from the intramodal 2L⇒2L scattering, conserving the total number $\int N_q^L dq$ of L-MEM quasiparticles, while $F_q^{LU}$ is the transport rate $N_q^L \to N_q^U$, caused by the intermodal 2L⇒2U scattering. Using a classical Hamiltonian approach, we estimated $\mu_q$ and $F_q^{LU}$ in the vicinity of the hybridization (see Appendix) as

$$\mu_q^L \simeq q_\times^3 (T_q^{LL})^2 (N_q^L)^3 \omega_\times^{-1},$$
$$F_q^{LU} \simeq q_\times^2 (T_q^{LU})^2 (N_q^L)^2 N_-^U \omega_\times^{-1}, \quad (4)$$

where $N_-^U$ is the density of U-MEMs at sufficiently large negative $o_q$, say at $o_q \simeq -5$. Using these estimates we rewrite the balance equation as

$$\frac{d}{dq}\left[\cos^8 \varphi_q (\mathcal{N}_q^L)^3\right] = 3a (\mathcal{N}_q^L)^2 \cos^4 \varphi_q \sin^4 \varphi_q. \quad (5)$$

Here, $\mathcal{N}_q^L = N_q^L/N_+^L$ is the dimensionless L-MEM density, normalized by the density $N_+^L$ at a sufficiently large positive $o_q$, say at $o_q \simeq +5$. The parameter $a = bN_-^U/N_+^L$ involves a dimensionless coefficient $b$ (presumably of the order of unity), which combines all the uncontrolled parameters in our estimates.

The ordinary differential equation (5) can be solved with the boundary condition $\lim_{q\to\infty} \mathcal{N}_q^L = 1$, giving the relative MEMs population in the hybridization region

$$\mathcal{N}_q^L = \frac{1}{\cos^{8/3}\varphi_q}\left[1 - a\int_q^\infty \frac{\sin^4\varphi_p dp}{\cos^{4/3}\varphi_p}\right]. \quad (6)$$

Solutions of Eq. (6) for different values of $a$, (Fig. 4f), exhibit a sharp rise in L-MEM population, caused by "*the bottleneck accumulation of MEM bosons near the hybridization region*". The maximum possible (asymptotic) bottleneck with an infinite growth of $\mathcal{N}_q^L \propto |q|^{-8/3}$ for large negative $o_q$ corresponds to the case where $a = 0$, in which only the intramodal scattering, conserving the total number of L-MEMs, is accounted for. For finite $a$, this growth is suppressed by the intermodal scattering that transfers quasiparticles from the L-MEM to the U-MEM. When this transfer is small ($a \ll 1$), the L-MEM density accumulation is quite significant, while for large $a \simeq 1$ (see two lower lines in Fig. 4f for $a = 1$ and 0.5) there is practically no MEM accumulation. Importantly, the increase in the population of the U-MEM during the course of the BEC formation ( i.e., increase in $N_-^U$ and, as a consequence, in $a$) must lead to the suppression of the bottleneck effect and to the consequent saturation of the MEM population peak. This effect is clearly visible in our observations shown in Fig. 3a.

The distance in the phase space between the BEC and MEM areas can be easily varied by the change of the bias magnetic field (see Figs. 2a,b), potentially allowing the control of the dynamics of magnon condensates and supercurrents [8] by the adjustment of their incoherent and coherent intermodal scattering with magneto-elastic bosons. It should be noted that according to our theory and in contrast to BEC, represented by a single macroscopic mode [8, 27, 28], the bottleneck accumulation occurs in a narrow but finite spectral region. The possibility of a coherent dynamics of accumulated quasiparticles constitutes an intriguing problem of fundamental importance for physics in general and for forthcoming magnonic applications.

The bottleneck accumulation phenomenon, reported here for a magnon-phonon gas, is not unique for this particular system and can occur in any multicomponent gas-mixture of interacting quasiparticles with significantly different scattering amplitudes. For example, the accumulation of magnetic polaritons [29] is expected in out-of-plane magnetized ferrimagnetic films.

Financial support from the Deutsche Forschungsgemeinschaft (project INST 161/544-3 within the Transregional Collaborative Research Centre SFB/TR 49 "Condensed Matter Systems with Variable Many-Body Interactions"), from EU-FET (Grant InSpin 612759) and from the State Fund for Fundamental Research of Ukraine (SFFR) is gratefully acknowledged. D.A.B. is supported by a fellowship of the Graduate School Material Sciences in Mainz (MAINZ) through DFG funding of the Excellence Initiative (GSC-266). As well we acknowledge I. I. Syvorotka (Scientific Research Company "Carat", Lviv, Ukraine) for supplying us with the YIG film sample.

A.A.S. and B.H. contributed to the experimental idea, planned and supervised the project. D.A.B., P.C., V.I.V., A.V.C., and A.A.S. carried out the experiments. D.A.B. contributed to the wavevector-resolved experimental set-up. V.S.L., G.A.M., and A.P. developed the theoretical model. A.P. carried out the numerical calculations. All authors analysed the experimental data and discussed the results.

## APPENDIX

**Sample and pumping circuit.** The spectral characteristics and temporal evolution of a magnon gas populated by parametric microwave pumping were studied in an in-plane magnetized $6.7\,\mu m$ thick single-crystal Yttrium Iron Garnet (YIG, $Y_3Fe_5O_{12}$) film grown in the (111) crystallographic plane on a Gadolinium Gallium Garnet (GGG, $Gd_3Ga_5O_{12}$) substrate by liquid-phase epitaxy. The magnon and phonon spectra presented in this Letter have been calculated using the saturation magnetization $4\pi M_s = 1750$ G, the exchange constant $\alpha = 3\cdot 10^{-12}$ cm$^2$ and the magneto-elastic constant $B_2 = 6.96 \cdot 10^6$ erg cm$^{-3}$. The pumping circuit (see Fig. 1a) is fed with $1.5\,\mu s$ long microwave pulses with a carrier frequency $\omega_p = 2\pi \cdot 13.62$ GHz. The pulses are applied with a repetition time of 1 ms allowing for magnetic and temperature equilibration of the sample between the pulses. A half-wave microstrip resonator with a width of $500\,\mu m$ (or $50\,\mu m$ in the test experiment with the narrow pumping area) creates an alternating Oersted field $h_p$ along the direction of the bias magnetic field $H$, realizing parallel parametric pumping [22]. The probing laser beam is focused onto a spot with a $25\,\mu m$ diameter on a YIG film placed in the middle of the resonator (see Fig. 1a).

**Time- and wavevector-resolved Brillouin light scattering spectroscopy.** The magnon and phonon dynamics were analysed by means of time- and wavevector-resolved BLS spectroscopy [30, 31] in back-scattering geometry. This spectroscopic technique is based on inelastic scattering of photons by quasiparticles of material waves. In the cause of the scattering process, a quasiparticle is either created (Stokes process) or annihilated (anti-Stokes process). As a result, the scattered photon acquires a frequency shift by the frequency of this quasiparticle and changes its wavevector due to the corresponding momentum transfer.

For example, in the case of a thin magnetic film, the in-plane component of the wavevector $q_L$ of the probing light is inverted by a counter-propagating magnon mode if the magnon wavenumber $q$ satisfies the momentum conservation condition $q = -2q_L \cos(\Theta)$, where $\Theta$ is the angle between $q_L$ and the film plane. By changing the angle $\Theta$, wavevector selection of in-plane magnons with wavevector $q$ can be implemented. In our experiment, the probing beam of 532 nm wavelength generated by a single-mode solid-state laser is focused onto a microstrip resonator placed just below the YIG film. In this case, the inelastically scattered light with the inverted in-plane wavevector component $-q_L \cos(\Theta)$ is reflected back to the objective and is collected for further analysis. Both the sample and the magnetic system are mounted on a rotating stage in order to be able to change the angle $\Theta = \Theta_{q_\parallel}$ (see Fig. 1), holding the magnetization conditions constant. By varying this angle, magnons with wavevectors parallel to the magnetic field $H$ are resolved [31]. The described setup allowed us to resolve wavevector values up to $1.5 \cdot 10^5$ rad/cm. For the particular experiment the wavevector resolution was $\pm 2 \cdot 10^3$ rad/cm.

The light collected by the objective is directed to a multipass tandem Fabry-Pérot interferometer [32–34] for frequency selection. At the output of the interferometer a single photon counting avalanche diode detector is placed. The output of the detector is connected to a fast data acquisition module, which enables accumulation and further analysis of the envelope of the temporal dynamics of the scattered photons. The time resolution is based on a stroboscopic technique – the experiment is repeated many times (the repetition rate of the experiment was 1 ms) with the same parameters. When the pumping pulse is applied, it triggers the internal counter in the acquisition module, which has an internal clock rate of 1 ns, defining the temporal resolution of the experimental setup [30]. Every time the detector registers a photon, this event is recorded to a database which collects the number of photons that arrived during each clock cycle. The frequency of the interferometer's transmission is also recorded, thus providing frequency information for each detected photon.

**Hybridized magnon-phonon Hamiltonian.** At room temperature we can restrict ourselves to the classical limit and describe the system of interacting magnons and phonons in the framework of a classical Hamiltonian formalism. This approach is applicable to a wide class of weakly interacting wave systems, allowing a physically transparent and very compact description of their common properties, see e.g. Chapter 1 in Ref. [35].

Introducing complex canonical amplitudes of magnons and phonons in the wave-vector $q$-representation, $a(q,t) \equiv a_q$ and $b(q,t) \equiv b_q$ we can write their Hamiltonian equation of motion as follows:

$$i\frac{\partial a_q}{\partial t} = \frac{\partial \mathcal{H}}{\partial a_q^*}, \quad i\frac{\partial b_q}{\partial t} = \frac{\partial \mathcal{H}}{\partial b_q^*}. \tag{7a}$$

Here the Hamiltonian function $\mathcal{H}$ (called hereafter for brevity "the Hamiltonian") is a functional of canonical variables $a_q$, $a_q^*$ for all $q$ and their complex conjugated counterparts $a_q^*$, $b_q^*$. We chose $\mathcal{H}$ as

$$\mathcal{H} = \mathcal{H}_2 + \mathcal{H}_4, \tag{7b}$$

$$\mathcal{H}_2 = \sum_q \left[ \omega_q^{\mathrm{m}} a_q a_q^* + \omega_q^{\mathrm{p}} b_q b_q^* + \frac{\Delta}{2}\left(a_q b_q^* + a_q^* b_q\right)\right], \tag{7c}$$

$$\mathcal{H}_4 = \frac{1}{4}\sum_{q_1+q_2=q_3+q_4} T_{12,34}\, a_1^* a_2^* a_3 a_4. \tag{7d}$$

Here, the first two terms in $\mathcal{H}_2$ describe the free propagation of magnons and phonons with the dispersion laws $\omega_q^{\mathrm{m}}$ and $\omega_q^{\mathrm{p}}$. The last two terms in $\mathcal{H}_2$ are responsible for their linear coupling due to magnetoelastic effect with a coupling amplitude $\Delta$. Nonlinearity in the acoustic system can be safely neglected in comparison with the strongly nonlinear spin-wave (magnon) subsystem.

The dispersion laws $\omega_q^{\mathrm{m}}$ and $\omega_q^{\mathrm{p}}$ cross at some hybridization wavenumber $q = q_\times$: $\omega_{q_\times}^{\mathrm{m}} = \omega_{q_\times}^{\mathrm{p}}$. We consider here a weak coupling regime $\Delta \ll \omega_{q_\times}^{\mathrm{m}} = \omega_{q_\times}^{\mathrm{p}}$ with a narrow hybridization region $\delta q$ around $q_\times$, such that $\delta q\, \partial(\omega_q^{\mathrm{p}} - \omega_q^{\mathrm{m}})/\partial q \simeq \Delta$.

Being interested in the system evolution in the hybridization region $\delta q$, we can restrict ourselves by the four-magnon interaction Hamiltonian $\mathcal{H}_4$, Eq. (7d) [35] with the sum restricted by the hyper-surface $q_1 + q_2 = q_3 + q_4$. The interaction amplitude $T_{12,34}$ depends on $q_1$, $q_2$, $q_3$, $q_4$; $a_j$ denotes $a(q_j,t)$ with $j = 1, 2, 3,$ and 4.

The quadratic Hamiltonian $\mathcal{H}_2$ can be diagonalized by the linear canonical Bogoliubov $(u,v)$-transformation of the form:

$$\begin{aligned} a_q &= c_q^{\mathrm{L}} \cos\varphi_q + c_q^{\mathrm{U}} \sin\varphi_q, \\ b_q &= -c_q^{\mathrm{L}} \sin\varphi_q + c_q^{\mathrm{U}} \cos\varphi_q. \end{aligned} \tag{8}$$

Choosing the coordinate system rotation angle $\varphi_q$ according to Eq. (1) we obtain a new diagonal Hamiltonian $\widetilde{\mathcal{H}}_2$ in terms of new normal canonical amplitudes of the U- and L-MEMs $c_q^{\mathrm{U}}$ end $c_q^{\mathrm{L}}$ with frequencies $\Omega_q^{\mathrm{U}} \equiv \Omega^+$ end $\Omega_q^{\mathrm{L}} \equiv \Omega^-$:

$$\widetilde{\mathcal{H}}_2 = \sum_q \left[\Omega_q^{\mathrm{U}} c_q^{\mathrm{U}} c_q^{\mathrm{U}*} + \Omega_q^{\mathrm{L}} c_q^{\mathrm{L}} c_q^{\mathrm{L}*}\right], \tag{9a}$$

$$\Omega_q^\pm = \frac{1}{2}\left\{\omega_q^{\mathrm{m}} + \omega_q^{\mathrm{p}} \pm \sqrt{[\omega_q^{\mathrm{m}} - \omega_q^{\mathrm{p}}]^2 + \Delta^2}\right\}. \tag{9b}$$

The interaction Hamiltonian (7d) in new variables includes terms $\mathcal{H}_4^{\mathrm{UU}}$, $\mathcal{H}_4^{\mathrm{LL}}$, responsible for the interactions within the upper- and lower-hybrid mode, respectively and a term $\mathcal{H}_4^{\mathrm{LU}}$, describing the interaction between them:

$$\mathcal{H}_4^{\mathrm{UU}} = \frac{1}{4}\sum_{q_1+q_2=q_3+q_4} T_{12,34}^{\mathrm{UU}} c_1^{\mathrm{U}*} c_2^{\mathrm{U}*} c_3^{\mathrm{U}} c_4^{\mathrm{U}},$$

$$T_{12,34}^{\mathrm{UU}} = T_{12,34}^{\mathrm{UU}} \sin\varphi_1 \sin\varphi_2 \sin\varphi_3 \sin\varphi_4, \tag{10a}$$

$$\mathcal{H}_4^{\mathrm{LL}} = \frac{1}{4}\sum_{q_1+q_2=q_3+q_4} T_{12,34}^{\mathrm{LL}} c_1^{\mathrm{L}*} c_2^{\mathrm{L}*} c_3^{\mathrm{L}} c_4^{\mathrm{L}},$$

$$T_{12,34}^{\mathrm{LL}} = T_{12,34}^{\mathrm{LL}} \cos\varphi_1 \cos\varphi_2 \cos\varphi_3 \cos\varphi_4, \tag{10b}$$

$$\mathcal{H}_4^{\mathrm{LU}} = \frac{1}{4}\sum_{q_1+q_2=q_3+q_4} T_{12,34}^{\mathrm{LU}}\left[c_1^{\mathrm{U}*} c_2^{\mathrm{U}*} c_3^{\mathrm{L}} c_4^{\mathrm{L}} + \mathrm{c.c.}\right],$$

$$T_{12,34}^{\mathrm{LU}} = T_{12,34}^{\mathrm{LU}} \sin\varphi_1 \sin\varphi_2 \cos\varphi_3 \cos\varphi_4. \tag{10c}$$

Here "c.c." stands for complex conjugation. The energy and particle fluxes within the upper and lower modes are proportional to $|T_{12,34}^{\mathrm{UU}}|^2$ and $|T_{12,34}^{\mathrm{LL}}|^2$, the energy and particle exchange between modes is proportional to $|T_{12,34}^{\mathrm{LU}}|^2$. To illustrate how these objects depend on the frequency near the hybridization crossover, where $\omega_q^{\mathrm{m}} = \omega_q^{\mathrm{p}}$, we plot them in Fig. 4b with the shorthand notations:

$$T_q^{\mathrm{UU}} \equiv T_{q,q;q,q}^{\mathrm{UU}}, \quad T_q^{\mathrm{LL}} \equiv T_{q,q;q,q}^{\mathrm{LL}}, \quad T_q^{\mathrm{LU}} \equiv T_{q,q;q,q}^{\mathrm{LU}}, \tag{11}$$

as functions of dimensionless distance to the crossover $o_q$ defined by Eq. (1).

**Statistical description of magneto-elastic modes.** Statistical description of weakly interacting waves can be obtained [36] in terms of a kinetic equation, shown below for the continuous limit, when the system size $\mathcal{L}$ is much larger than the length $2\pi/q$:

$$\frac{\partial n^{\mathrm{L}}(q,t)}{\partial t} = \mathrm{St}^{\mathrm{L}}(q,t), \quad \frac{\partial n^{\mathrm{U}}(q,t)}{\partial t} = \mathrm{St}^{\mathrm{U}}(q,t). \tag{12}$$

Here $n^{\mathrm{L}}(q,t) \equiv n_q^{\mathrm{L}}$ and $n^{\mathrm{U}}(q,t) \equiv n_q^{\mathrm{U}}$ are the simultaneous pair correlations of the L- and U-MEMs, defined by the following expressions: $\langle c_q^{\mathrm{L}} c_{q'}^{\mathrm{L}}\rangle = \frac{4\pi^2}{\mathcal{L}^2}\delta(q+q') n_q^{\mathrm{L}}$, $\langle c_q^{\mathrm{L}} c_{q'}^{\mathrm{L}}\rangle = \frac{4\pi^2}{\mathcal{L}^2}\delta(q+q') n_q^{\mathrm{L}}$, where $\langle\ldots\rangle$ stands for the ensemble averaging. In the classical limit, the dimensionless ration $n_q(q,t)/\hbar$ are the quantum mechanical occupation numbers of Bose particles. Here $2\pi\hbar$ is the Plank constant. In what follows we discuss only relevant L-MEM population $n_q^{\mathrm{L}}$.

The collision integral $\mathrm{St}^{\mathrm{L}}(q,t)$ may be found in various ways [36], including the Golden Rule, widely used in quantum mechanics. Accounting for the 2L⇒2L and 2L⇒2U scattering, we have

$$\mathrm{St}^{\mathrm{L}}(q,t) = \mathrm{St}^{\mathrm{LL}}(q,t) + \mathrm{St}^{\mathrm{LU}}(q,t), \tag{13a}$$

$$\mathrm{St}^{\mathrm{LL}}(q,t) = \frac{\pi}{4}\int dq_1 dq_2 dq_3\, \delta(q+q_1-q_2-q_3) \tag{13b}$$
$$\times \delta(\Omega_q^{\mathrm{L}} + \Omega_1^{\mathrm{L}} - \Omega_2^{\mathrm{L}} - \Omega_3^{\mathrm{L}}) |T_{q1\,23}^{\mathrm{LL}}|^2$$
$$\times [n_2^{\mathrm{L}} n_3^{\mathrm{L}}(n_q^{\mathrm{L}} + n_1^{\mathrm{L}}) - n_q^{\mathrm{L}} n_1^{\mathrm{L}}(n_2^{\mathrm{L}} + n_3^{\mathrm{L}})],$$

$$\mathrm{St}^{\mathrm{LU}}(q,t) = \frac{\pi}{4}\int dq_1 dq_2 dq_3\, \delta(q+q_1-q_2-q_3) \tag{13c}$$
$$\times \delta(\Omega_q^{\mathrm{U}} + \Omega_1^{\mathrm{U}} - \Omega_2^{\mathrm{L}} - \Omega_3^{\mathrm{L}}) |T_{q1\,23}^{\mathrm{LU}}|^2$$
$$\times [n_2^{\mathrm{U}} n_3^{\mathrm{U}}(n_q^{\mathrm{L}} + n_1^{\mathrm{L}}) - n_q^{\mathrm{L}} n_1^{\mathrm{L}}(n_2^{\mathrm{U}} + n_3^{\mathrm{U}})].$$

**Estimates of the flux and the L→U transfer rate.** Assuming for simplicity isotropy of the problem and introducing a 1D version of the L-MEM density $N_q^{\rm L} = 4\pi q^2 n_q^{\rm L}$, we can present the kinetic equation (12) as follows

$$\frac{\partial N_q^{\rm L}}{\partial t} = 4\pi q^2 \left[ {\rm St}^{\rm LL}(q,t) + {\rm St}^{\rm LU}(q,t) \right]. \qquad (14)$$

The collision term ${\rm St}^{\rm LL}(q,t)$ preserves the total number of the L-MEMs and thus may be represented in a divergent form $d\mu_q^{\rm L}/dq$, where $\mu_q$ is the L-MEM quasiparticle flux towards small wavenumbers. Together with Eq. (3) this gives $\mu_q = 4\pi \int^q \tilde{q}^2 {\rm St}^{\rm LL}(\tilde{q}) d\tilde{q}$. Now, using Eq. (13b), we obtain the estimate (4) for the flux $\mu_q^{\rm L}$.
We also need to account for the LL⇒UU scattering described by the collision integral ${\rm St}^{\rm LU}$ [Eq. (13c)]. Since, in the crossover region we expect $N_q^{\rm L} > N_q^{\rm U}$, the leading contribution to ${\rm St}^{\rm LU}(q) \propto (N_q^{\rm L})^2 N_q^{\rm U}$ is the negative term:

$${\rm St}^{\rm LU}(q) \simeq -(T_q^{\rm LU})^2 (N_q^{\rm L})^2 N_q^{\rm U} \omega_\times^{-1}. \qquad (15)$$

Comparing Eqs. (3) and (14) we conclude that $F^{\rm LU} = -4\pi q^2 {\rm St}^{\rm LU}(q)$. Together with Eq. (15) this gives the estimate (4) for the transfer rate $F_q^{\rm LU}$ of L-MEM quasiparticles to U-MEM quasiparticles.